\documentstyle[12pt,epsfig]{article}
\author{Diego Harari, Silvia Mollerach and Esteban Roulet\\
CONICET, Centro At\'omico Bariloche, \\
Av. Bustillo 9500, Bariloche, 8400, Argentina.}

\title{Statistics of cluster multiplicity and the nature of UHE cosmic ray sources}

\begin{document}

\maketitle

\begin{abstract}
We study in detail the properties of clusters of ultra high energy 
cosmic ray events, 
looking in particular to their angular correlation function, to the relative 
frequency of clusters with different multiplicities (e.g. doublets vs. triplets
or quadruplets), as well as the way in which these quantities should
evolve for different ultra high energy cosmic rays source scenarios as a 
function of the experimental exposure achieved.
We identify some useful tools which can help to characterise the nature of
the cosmic ray sources in a more precise way 
after a modest increase in statistics will be achieved 
in the very near future, even before strong signals from individual sources
become eventually observable.

\end{abstract}

\section{Introduction}

In spite of the big effort done over the last decades to unravel the origin 
and nature of the cosmic rays (CRs) at ultra high energies (UHE), most of 
the relevant questions related to them remain still unanswered 
\cite{cr04}. Certainly 
the main obstacle has been the scarcity of data, which is due
to the smallness of the fluxes at the highest energies
(e.g. $\sim 1/{\rm km}^2$yr for $E>10^{19}$~eV). The observed isotropy 
of CRs above the
ankle (i.e. for $E>5\times 10^{18}$~eV), suggests that at these energies 
CRs are of extragalactic origin, but the nature of these particles (whether 
they are protons, heavier nuclei, photons or even neutrinos with stronger 
interactions), the nature of their sources (e.g. steady ones like active galactic nuclei
jets, transient ones like gamma ray burst (GRBs) or diffuse ones like decaying topological
defects) and the mechanism responsible for their acceleration in bottom-up 
scenarios or production in top-down ones (and hence the original source spectra) 
are all under strong debate. All these questions are further challenged by the 
(lack of) observation of the theoretically expected GZK suppression above 
$\sim 6\times 10^{19}$~eV, due to the attenuation that CR protons 
coming from far away sources should suffer due to their interactions
with CMB photons (or similarly due to the photodisintegration processes that
would affect the propagation of heavier nuclei). Clearly a big boost
in our understanding of these issues could be obtained if individual
CR sources were identified, and this search is indeed the main purpose
of the next generation of observatories, like the Pierre Auger one now
under completion.

There has been a lot of discussion during the last years about possible hints 
in the existing data of an excess of clustering in 
the highest energy range ($E>4\times 10^{19}$~eV), that could be already 
indicating the possible emergence of the first strong CR sources out of
a more diffuse background of unresolved sources. This excess clustering (with
respect to the expectations from chance coincidences of an isotropic distribution)
has been observed by the AGASA group \cite{agasa}(which detected 4 doublets and one triplet
within 2.5$^\circ$, out of a total of 57 events\footnote{We are not including 
in our discussion one event with energy slightly below $4\times 10^{19}$~eV, 
included by AGASA because it lead to an extra doublet.}) and also in a 
combined analysis
of data from several experiments \cite{uc00} (Yakutsk, Haverah Park, AGASA and Volcano Ranch, 
which obtained 8 doublets and 2 triplets within $4^\circ$ out of a total of 92 events),
although it does not seem to be present in Fly's Eye or Hi Res data \cite{hires}.

In this work we will study in detail the properties of the event clusters
expected in different scenarios of UHECR sources, with the aim of devising
the most appropriate tools to discriminate among them. The results 
we obtain suggest that a modest increase in the present statistics (as
could be achieved after one year of AUGER observations) 
will allow to clarify significantly the existing debate, even before
any strong (and hence unambiguous) signal from an individual CR source
is observed.

\section{Clusters from chance coincidences}

Before considering the possibility that the clusters are due to strong CR 
sources with large enough fluxes so as to have induced more than one event
with the present experimental exposures, it is important to consider the expected
clustering signal from just chance coincidences in the limit in which
all events observed were produced by independent sources, which are too 
weak to produce by themselves an individual cluster (i.e., considering that
there is a very large number of sources, all with a small chance of
producing even a single event in the present experiments). As we show below, 
this probability depends strongly on the way the sources are distributed
across the sky, and so we start with the simplest case of an isotropic
distribution.

\subsection{Isotropic sources}

A commonly adopted measure of a clustering signal is obtained from the study
of the angular correlation of pairs, d$N_p/$d$\theta$, which gives the
number of event pairs with angular separation between $\theta$ and $\theta$
+ ${\rm d}\theta$ expected from a total number $n$ of events. 
In the case of an isotropic distribution, it can be written as

\begin{equation}
{{\rm d}N_p\over {\rm d}\theta}\propto \sin\theta\int {\rm d}\Omega_1
\int {\rm d}\Omega_2\ \omega(\delta_1)\omega(\delta_2) \ \delta(\hat\Omega_1
\cdot\hat\Omega_2-\cos\theta),
\end{equation}
where d$\Omega_i=\cos\delta_i$d$\delta_i$d$\alpha_i$ in terms of declination
and right ascension, and $\delta_i$ has to be integrated over the declinations 
observable to the experiment.  
The normalization is such that the total number of pairs 
(i.e. integrating over $\theta$) is just $n(n-1)/2$.

The function $\omega(\delta)$ takes into account the declination dependence 
of the exposure of the experiment under consideration, and for a 
surface detector\footnote{For the air-fluorescence technique the exposure is more
uniform in declination, but has also a dependence on right ascension (see e.g. 
\cite{hires}).} is given by \cite{so01}
\begin{equation}
\omega(\delta)\propto \cos b_0\cos\delta \sin\alpha_{m}+\alpha_m \sin b_0
\sin\delta,
\end{equation}
where $b_0$ is the latitude of the experiment (e.g. $b_0=35.8^\circ$ for AGASA) and
\begin{equation}
\alpha_m=\left\{ \matrix{0\cr {\rm acos}\ \xi \cr
\pi}\right.\ \ \ 
\matrix{\xi>1\cr  -1\leq \xi\leq 1\cr \xi<-1}
\ \ \ \ \xi={\cos\theta_m-\sin b_0\sin\delta\over 
\cos b_0\cos\delta},
\end{equation}
with $\theta_m$ the maximum zenith angle of the CR showers considered ($\theta_m=
45^\circ$ for the AGASA sample).

\begin{figure}[t]
\centerline{{\epsfig{width=4.in,file=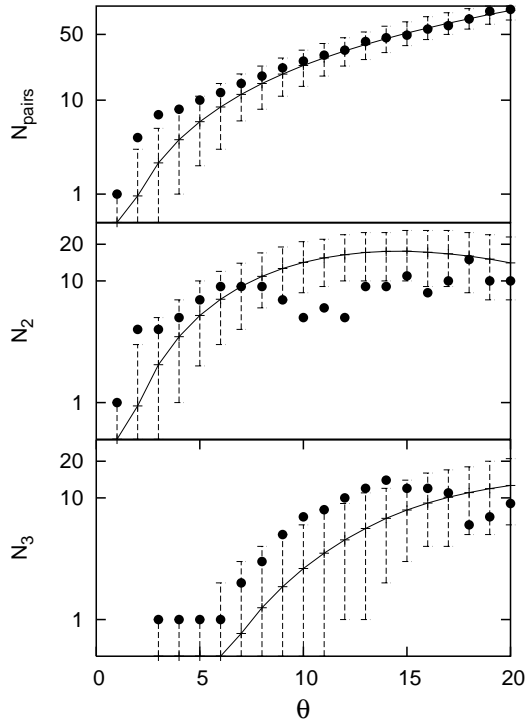,angle=-90}}}
\caption{Number of pairs (upper panel), doublets (middle) and triplets (lower)
separated by an angle smaller or equal than $\theta$. The solid
lines follow the mean of 10000 random realizations of 57 events isotropically distributed
with AGASA's exposure. Error bars contain 95\% of the results
in the simulations. Points correspond to AGASA data above $4\times 10^{19}$ eV.
Significant departures of the data from random expectations are apparent at angular scales smaller than a few degrees in all three plots, as 
well as at scales between 9 and $14^\circ$ in the distribution of doublets and triplets
(but not in the number of pairs).} 
\label{pdt}
\end{figure}

Potential point sources of UHECRs should manifest themselves through a significant
enhancement above random expectations of the correlation function at values 
around and below the angular reconstruction accuracy. Energy-dependent deflections 
of CR trajectories by intervening magnetic fields may disperse the arrival directions,
and thus clustering signals may be spread over larger angular scales. It thus proves
convenient to analyse the integrated correlation function $N_p(<\theta)=\int_0^\theta 
{\rm d}\theta^\prime~{\rm d}N_p/ {\rm d}\theta^\prime$, which is 
simply the total number of event pairs which are separated by an angle smaller 
than $\theta$ (albeit the most common practice is to look instead at the 
distribution ${\rm d}N_p/{\rm d}\theta$). 
The number of pairs at angles smaller than $\theta$ is 
plotted in the upper panel in Figure~\ref{pdt}, for a total number of 57 events 
and for the AGASA exposure, and is compared  
with the AGASA data. The solid line follows the mean value of 10000 simulated 
isotropic random realizations. The error bars contain 95\% of the results in the
random realizations at each angular scale (the larger and smaller 2.5\% portions 
were discarded). The points are the AGASA data. An excess of pairs
with respect to random expectations is apparent at small angles ($\theta<$ 
few degrees), which is related to the observation of 4 doublets and 
one triplet (corresponding to three additional pairs) at separations smaller 
than 2.5$^\circ$ in the AGASA data, while on average 1.45 doublets
would be expected. The probability \footnote{The significance of these probabilities is 
a matter of debate~\cite{ti01,ev03,fi03}, given its dependence on the size of the angular 
bin and the energy threshold.} 
of getting seven or more pairs at 
$\theta<2.5^\circ$ in the random realizations is $\sim 1.5\times 10^{-3}$.

The angular correlation of pairs is the lowest order of a hierarchy of correlation
functions. Additional information on the clustering process is encoded, for instance,
in the distribution of clusters with different multiplicities. The middle and lower panels 
in Figure~\ref{pdt} plot respectively the mean number of doublets, $N_2(<\theta)$, and triplets, 
$N_3(<\theta)$, in random realizations of 57 events with the AGASA exposure, compared with the AGASA data.  
The excess clustering at small angular scales in AGASA data is noticeable. 
The probability of forming four doublets at angular scales below $2.5^\circ$ is 
$\sim 0.06$, and the probability of forming one triplet by chance is $\sim 0.016$. 
The probability of forming 4 doublets and one triplet is of course even smaller 
than that of forming seven pairs, and is of order $\sim 9\times 10^{-4}$.

Notice that while the number of event pairs in AGASA data shows no significant departures 
from isotropic expectations at angular scales larger than a few degrees, their distribution
in clusters of different multiplicities deviates noticeably from the mean at
angular scales roughly between $9^\circ$ and $14^\circ$.  The number
of doublets is in defect and the number of triplets is in excess with respect to  
random isotropic realizations, at a probability level comparable to the small angular scale
clustering signal. Spreading of arrival directions from point UHECR sources by
magnetic fields
and structure in their spatial distribution are among conceivable causes
of such departures. We will come back to a more thorough characterization of this 
signal in section~4, with alternative statistical tools.

One can get a feeling about the amount of clusters expected for different 
multiplicities by considering the following simplified model to describe
the way clusters are produced by chance coincidences of random isotropic data (a
similar approach was adopted in ref.~\cite{go01} to obtain the probabilities
for observations of different numbers of clusters).
Let us first ignore the possible effects of a non-uniform exposure, 
and let us divide the total solid angle $\Omega$ monitored by the experiment
($\Omega\simeq 7.2$~sr for AGASA when $\theta_{max}=45^\circ$ is considered) into
$N$ solid angle bins of characteristic size $\Delta \Omega\simeq \pi \theta^2$,
so that $N\simeq \Omega/\Delta\Omega
\simeq 1050\ (\Omega/2\pi)(2.5^\circ/\theta)^2$ for small angles
\footnote{$N=(\Omega/2\pi)/(1-\cos\theta)$ when $\theta$ is not small enough.}.
Now, we start adding events
up to a total number $n$ ($n=57$ for the published AGASA data to which we 
want to compare), assuming that each event has the same probability ($p=1/N$) 
to fall into any of the independent solid angle bins. When two events
fall into the same bin, this will then produce a doublet with characteristic
separation smaller than $\theta$. It is then simple to show that the average 
number of clusters $\overline{N}_m(n)$ with a given multiplicity $m$ 
(e.g. $m=2$ for doublets, etc.) after observing $n$ events satisfy
\begin{equation}
\overline{N}_m(n+1)=\left(1-p\right)\overline{N}_m(n)+p\ \overline{N}_{m-1}(n)
\end{equation}
and hence they are just given by a binomial distribution
\begin{equation}
\overline{N}_m(n)=\left(\matrix{n\cr m}\right)p^{m-1}(1-p)^{n-m}.
\label{bin}
\end{equation}
In particular, one has
\begin{equation}
\overline{N}_1(n)=n\left(1-\frac{1}{N}\right)^{n-1}\simeq 
n\ {\rm exp}(-n/N)
\end{equation}
\begin{equation}
\overline{N}_2(n)={n(n-1)\over 2N}\left(1-\frac{1}{N}\right)^{n-2}\simeq 
\frac{n^2}{2N}\ {\rm exp}(-n/N)
\end{equation}
\begin{equation}
\overline{N}_3(n)={n(n-1)(n-2)\over 6N^2}\left(1-\frac{1}{N}\right)^{n-3}\simeq 
\frac{n^3}{6N^2}\ {\rm exp}(-n/N)
\end{equation}
where the right-most expressions hold for $n\gg 1$.
The exponential factors are actually only relevant when the total 
number of events becomes of the order of the number of available solid 
angle bins, since in this case most of the events appear in clusters and 
the multiplicity of the bins rapidly grows suppressing the number of 
low multiplicity clusters. Clearly by the time this will happen at small
angular scales, the 
discussion here will be redundant since already the statistics would have 
become
large enough to allow the origin of the observed clustering to be clarified, 
and hence 
we will focus the discussion here mainly to the situation in which 
$n\ll N$. In this 
case we see that the number of pairs grows quadratically with the number
of events observed (and hence with the experimental exposure), the number
of triplets grows as the third power and so on, as expected.
On the other hand, the ratio between the number of triplets and doublets
is $\overline{N}_3(n)/\overline{N}_2(n)\simeq n/3N$, 
and is hence very small as long as $n\ll N$, 
as should be the case at present for AGASA.

\begin{figure}[t]
\centerline{{\epsfig{width=4.in,file=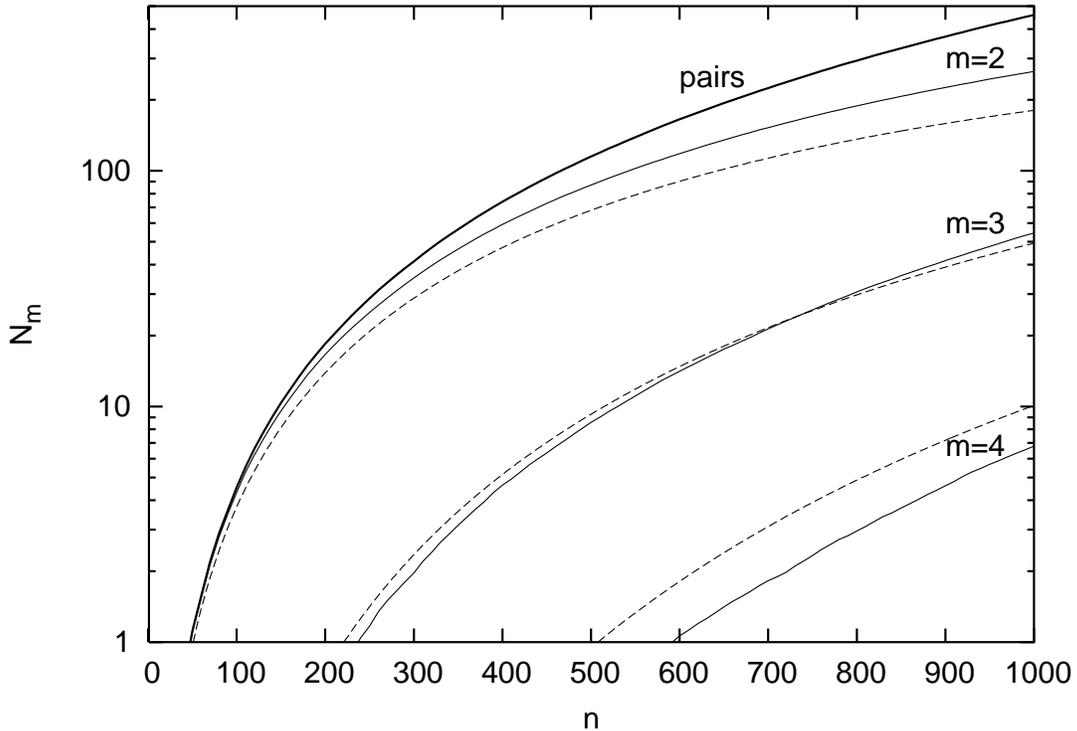,angle=-90}}}
\caption{Mean number of pairs (thick solid line), doublets, triplets and quadruplets
(thin solid lines) in simulated realizations of $n$ events randomly distributed with
AGASA's declination-dependent exposure. The result of the simple analytic model of 
eq.~(\ref{bin}) for doublets (m=2), triplets (m=3) and quadruplets (m=4) is plotted 
with dashed lines.} 
\label{Nn}
\end{figure}

Hence, a generic prediction of clustering at small angular scales
by chance coincidences of an 
isotropic distribution is that one expects first to see a growing number
of doublets
($\propto {\cal E}^2$, where ${\cal E}$ is the exposure achieved, and of
course one expects $n\propto {\cal E}$) with no  clusters of higher 
multiplicities (except for very unlikely 
fluctuations). Then, when the total number
of events becomes $\sim (6N^2)^{1/3}$, triplets start to appear with a faster
rate of growth ($\propto {\cal E}^3$). When this happens the number of 
doublets is typically $\overline{N}_2\simeq N^{1/3}$, 
which is already not a small number.
We see then that it is not only surprising to have 4 doublets in the AGASA data
when 1.5 ($\sim 57^2/2000$) were expected, but it is probably even more 
surprising to have observed a triplet out of only 57 events. As an example, 
we show in Figure~\ref{Nn} the average results obtained from many random 
isotropic realizations, taking 
into account a declination dependent exposure like that of AGASA, and plot 
the resulting number
of pairs, doublets, triplets and quadruplets obtained (for $\theta<2.5^\circ$) as
a function of the number of events (solid lines). The
main features just mentioned that result from the simple model (dashed lines)
are clearly apparent.

The simple model predicts a smaller
number of doublets with respect to the Monte Carlo results, 
what can be understood
because the model with solid angle bins 
requires that each event be asigned to a unique cluster, and thus 
does not include properly all possible configurations. Consider as an
example the case of
three events aligned that lead to two doublets but not a triplet
\footnote{Alignment of events can be caused by energy-dependent deflections in
magnetic fields, which makes even more relevant to properly discriminate such
a clustering signal.}.
In this case, one of the events participates in two different doublets, but 
clearly no event can fall in two different angular bins simultaneously.
The simple model also predicts an excess (at relatively small $n$) of
triplets and quadruplets with respect to the Montecarlo results. This can be
understood because once two events are separated by an angle $<\theta$, to form a 
triplet the third event has to fall within a solid angle smaller, on average,
than $\pi\theta^2$. Hence, the effective number of bins that define the probability
of higher multiplicity clusters is larger than the value of $N$
used to calculate the probability of forming doublets. At relatively larger $n$,
when clusters of multiplicity $m+1$ start to form, the simple model predicts instead
a defect of clusters of multiplicity $m$, for reasons analogous to those that
explain the defect in doublets. 
Notice that this should also affect the probabilities 
obtained in \cite{go01}.
The limitations of the simple model appear at
smaller values of the number of events $n$ when larger angular separations are
considered, since the number of bins $N$ becomes significantly smaller.

\begin{figure}[t]
\centerline{{\epsfig{width=4.in,file=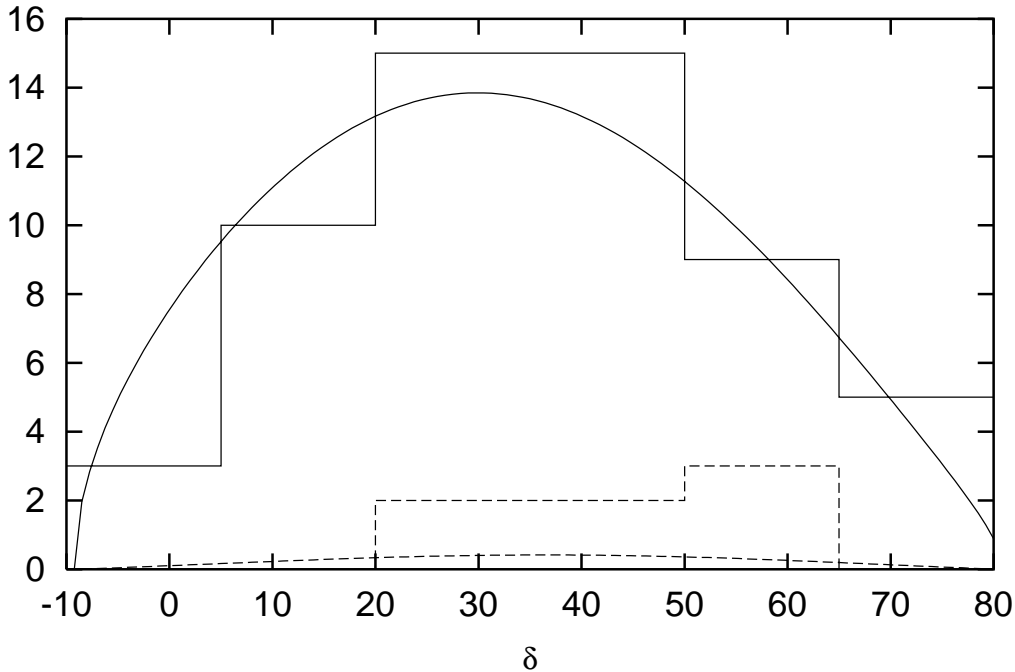,angle=-90}}}
\caption{Distribution in declination of AGASA events (solid histogram) and AGASA event pairs
(dashed histogram) compared to that expected from chance coincidences of an isotropic
distribution with AGASA's declination dependence of the exposure (solid and dashed curves).} 
\label{histo1}
\end{figure}

Another property of the multiplets that result from chance coincidences 
of an isotropic distribution is that, once we take into account that the 
relative exposure is proportional to $\omega(\delta)$, and hence that the number
of observed events should follow d$n/$d$\delta\propto \omega(\delta)\cos\delta$
(with the $\cos\delta$ factor just from the solid angle associated to $\delta$), 
the distribution of event pairs on the sky should be proportional to 
d$N_p\propto \omega(\delta)^2\cos\delta$d$\delta$,
and similarly the number of triplets (including those in quadruplets and 
higher multiplets) should behave as 
d$N_t\propto \omega(\delta)^3\cos\delta$d$\delta$, and so on. Hence, 
the high multiplicity clusters should appear more significantly concentrated 
at declinations close to the latitude of the experiment, where 
$\omega(\delta)$ is maximal.
This also implies that taking into account the declination dependence 
of the exposure will slightly enhance the ratio $\overline{N}_3/\overline{N}_2$
 (as well as 
the value of $\overline{N}_2$ itself) with respect to the estimates done neglecting
$\omega(\delta)$. We show in Figure~\ref{histo1} the observed distribution (solid histogram) 
of events and of event pairs (dashed histogram) in AGASA data, together with the 
expectations from chance 
coincidences out of an isotropic distribution.
It is  clear that although the angular distributions are consistent 
with the hypothesis of isotropic sources, there is an overall excess
in the number of pairs with respect to expectations from chance coincidences.

\subsection{Structured sources}

As is well known, the matter distribution in the local Universe is 
rather inhomogeneous, specially if we consider the neighbourhood where
trans-GZK events may originate (distances $<100$~Mpc). In this region indeed 
large voids are found, and most of the galaxies are distributed on large 
scale structures like the great attractor, the supergalactic plane, 
etc., which only cover a relatively small fraction of the sky (see e.g.
Figure~10 in \cite{cr04}). If UHECR 
sources were to follow a similar distribution, this can have a significant 
impact on the statistics of the clustering produced by chance overlaps
among different sources. For instance, if the sources were distributed 
uniformly over the sky but `avoiding voids', so that the effective solid 
angle where sources can be found is smaller than $\Omega$, one could repeat
the reasoning of the previous section and find similar expressions but
now with a smaller number $N'$ of solid angle bins where clusters could
be formed. More generally, if we substitute in the expressions 
of the previous section $n/\Omega\equiv \rho$, which would be the 
density of events per unit solid angle, it is clear that in the 
limit in which $\rho\Delta\Omega\ll 1$ the generalization of the expression for
the number of doublets for the case in which the density of sources is
not uniform becomes
\begin{equation}
\overline{N}_2(n)\simeq {\Delta\Omega\over 2}\int {\rm d}\Omega \rho^2
\end{equation}
If we now define the average density of events as
\begin{equation}
\overline\rho=\frac{1}{\Omega}\int {\rm d}\Omega\rho
\end{equation}
we obtain
\begin{equation}
\overline{N}_2(n)\simeq \overline{N}_2^{iso}(n)
\langle\left(\rho/\overline \rho\right)^2\rangle,
\end{equation}
where
\begin{equation}
\overline{N}_2^{iso}(n)=\frac{\Delta\Omega}{2}\overline\rho^2\Omega
\end{equation}
is the average number of doublets expected from an isotropic distribution 
(the $\overline{N}_2$ computed in the previous section), while
\begin{equation}
\langle\left(\rho/\overline \rho\right)^2\rangle\equiv
\frac{1}{\Omega}\int{\rm d}\Omega \left(\rho/\overline \rho\right)^2
\end{equation}
is a quantity bigger than unity.
If one assumes that the distribution of UHECR sources somehow follows the 
distribution of visible matter, one expects that  if the sources are 
indeed nearby (at least for trans-GZK events) there should be 
a significant non-uniformity
in the arrival directions of the observed events, probably leaving void 
regions in the sky and with the events finally falling, once larger 
statistics will be attained, along large scale structures reminiscent 
to strips across the sky (at these energies, deflections due to magnetic 
fields are not expected to be large, at least for CR protons). Notice that
due to the limited statistics accumulated so far, such a picture is 
not inconsistent with current observations.

As a simple example, suppose that only a fraction $\xi$ of the
observed sky has a uniform distribution of sources, while the rest consists
of void regions. This would lead to $\rho=\overline\rho/\xi$ in the regions 
with sources and hence $\langle\left(\rho/\overline \rho\right)^2\rangle=
1/\xi$, which shows that a significant enhancement in the number of pairs
can be achieved by distributing the sources anisotropically. This is understood
since in the higher density regions the probability of forming a doublet
by chance coincidences is largely enhanced ($\propto \rho^2$). An enhancement 
in the number of doublets 
 was indeed reported in numerical simulations with a structured 
source distribution \cite{me01}.
An even more pronounced effect should also result for the probability
of formation of triplets, which can similarly be shown
to behave as 
\begin{equation}
\overline{N}_3(n)\simeq 
\overline{N}_3^{iso}(n)\langle\left(\rho/\overline \rho\right)^3\rangle.
\end{equation}

\section{Clustering from individual sources}

The alternative explanation that has been suggested for the observed 
clustering is that the 
events in each multiplet originated in the same sources, 
so that the doublets and triplets observed would correspond
to the first manifestations of the brightest sources in the sky. In this
scenario, the clusters produced by chance coincidences discussed in the previous
section would just be a background for the signal searched, and the excess
observed should be attributed to individual sources.

Given a source with a certain flux $F$ (for energies above some 
specified threshold), the probability that it gives rise to a certain 
number $m$ of events follows a Poissonian  distribution \cite{wa97}
\begin{equation}
P_m(F)={\overline m^m\over m !}\ {\rm e}^{-\overline m}
\end{equation}
where the mean multiplicity is $\overline m=F{\cal E}$, with ${\cal E}$ being the 
experimental exposure towards the direction of the source (for definiteness we will
assume that the exposure is uniform over the whole sky, and just comment 
later on the effects of non-uniform exposures).
Hence, if d$n_s/$d$F$ denotes the number density of sources leading to a flux
between $F$ and $F+$d$F$, the average number of multiplets expected will be \cite{du00}
\begin{equation}
\overline N_m=\int_0^\infty {\rm d}F\ {{\rm d} n_s\over {\rm d}F}P_m(F)
\label{poisson}
\end{equation}

On the other hand, the probability of finding $k$ clusters of multiplicity $m$
will also be Poisson distributed, i.e. 
\begin{equation}
P_m(k)={\overline N_m^k\over k !}\ {\rm e}^{-\overline N_m}.
\label{poisson2}
\end{equation}

Regarding the source distribution function, the simplest situation would be
to assume a uniform distribution of sources with similar intrinsic
CR luminosities $L$. In this case, if we further neglect for simplicity 
the effects of energy losses, the  sources leading to a flux
larger than $F$ would just be those at a distance closer than $d=\sqrt{L/4\pi F}$,
what gives
\begin{equation}
n_s(>F)=\frac{4}{3}\pi \rho_s\left({L\over 4\pi F}\right)^{3/2},
\end{equation}
with $\rho_s$ the source volume density.
Hence, in this case one finds
\begin{equation}
{{\rm d}n_s\over {\rm d}F}\propto F^{-5/2}.
\end{equation}

Notice that this power law behaviour would lead to a divergent total flux 
if extrapolated down to very small fluxes (this is nothing else than 
Olber's paradox), and it is clear that CR attenuation during propagation as 
well as the effects of the universe expansion and source evolution will 
ultimately smooth out the faint end of the source distribution. The simplest 
way to cure this is to just introduce a lower cutoff flux 
$F_{min}$ below which no sources contribute, and since we are mainly interested 
on the effects of the bright end of the distribution, our main conclusions will be
independent of the particular way in which this cutoff is handled.

Another simple scenario which has been considered is that the CR sources are 
transient, such as in the models of acceleration during GRBs.
In this case, due to the energy dependent time delays $\tau(E,d)$ 
suffered by CRs after traversing a distance $d$ across a magnetised universe, 
a bursting 
source will be seen (neglecting energy losses and strong lensing effects)
as a monochromatic flux at the observer, with an energy  slowly decreasing 
with time but on a scale typically much longer than the duration of the 
experiments \cite{mi96}. However, stochastic energy losses during CR propagation 
introduce a spread in time delays $\Delta\tau(d,E)$ which leads to the
observation of CRs with a certain spread of energies at any given time, 
with $\Delta E/E\sim \Delta\tau/\tau$. Moreover, when strong magnetic lensing
effects are present, so that a CR source is actually observed as a 
very large number of multiple images associated to different alternative paths
from the source to the observer, this also introduces a spread in the 
energies of the particles observed at any given time. There are two ways
in which these processes affect the number of sources above a given 
flux\footnote{For transient sources, the flux should be considered  
in a given energy bin around $E$ \cite{mi96}.}
which can be observed at any given moment. First,  the longer is the 
spread in time delays, the more likely it becomes for a source to contribute
to the CR fluxes during the period of  observation, but however since the 
CRs produced in the burst will spread over a longer time interval, the flux will
be more suppressed than the naive $d^{-2}$ behaviour. As argued in 
ref.~\cite{mi96}, one expects $\Delta\tau\propto \tau\propto(d/E)^2$, and
hence one has that $F\propto 1/(d^2\Delta\tau)\propto d^{-4}$. Hence, 
the number of sources above a certain flux (i.e. within an associated distance
$d\propto F^{-1/4}$) will be \cite{mi96}
\begin{equation}
n_s(>F,E)\propto \frac{4}{3}\pi d^3\ \Delta\tau \propto F^{-5/4}.
\end{equation}
This gives a differential source density
\begin{equation}
{{\rm d}n_s\over {\rm d}F}\propto F^{-9/4}.
\end{equation}
In view of the previous results, we may then parametrise  the source distribution
with a simple power law
\begin{equation}
{{\rm d}n_s\over {\rm d}F}={\alpha N_s\over F_{min}}\left({F_{min}\over F}
\right)^{\alpha+1},
\end{equation}
with a lower cutoff at the minimum flux $F_{min}$ and where $N_s$ is the 
total number of sources above $F_{min}$. For the steady source model
one has $\alpha=3/2$ while for the bursting scenario $\alpha=5/4$.
For these simple scenarios one obtains, from eqs.~(\ref{poisson}) and 
(\ref{poisson2}), an expected average number
of multiplets 
\begin{equation}
\overline N_m={\alpha N_s\over m !}\Gamma(m-\alpha,
F_{min}{\cal E})\ (F_{min}{\cal E})^\alpha 
\end{equation}
with $\Gamma(x,y)$ the incomplete Gamma function.
Since we are interested in cases where $1<\alpha<2$, one finds that as long as
${\cal E}\ll F_{min}^{-1}$ (i.e. that the faint sources have  a small 
probability of leading to one event) 
the expected average  number of event clusters (with 
$m\ge 2$) will have very little sensitivity to the 
second argument of the Gamma function, leading to
$ \overline{N}_m\propto {\cal E}^\alpha {\Gamma(m-\alpha)/ m !}$,
while the number of single events (not in clusters) 
does depend on the faint end of the source
distribution\footnote{Since our treatment of the faint end has been rather 
crude, one should not use statistics such as the number of single vs. double 
events to get information on the number of sources contributing, and in this 
respect the parameter $N_s$ just comes out as a normalisation constant. 
An estimate of 
the number of sources producing CRs depends sensitively on the way the faint 
end is handled, and will require other general features  of the CR sources, 
such as the luminosity distribution, to be understood as well 
\cite{du00,fo01,is02,bl04}.}, having the 
approximate behaviour  $\overline N_1\simeq
\alpha N_s{\cal E}F_{min}/(\alpha-1)$. Clearly these approximate behaviours 
cease to be valid when ${\cal E}\simeq F_{min}^{-1}$, but before that happens
(as should be anyway the case for the next few years) one has that while
the average number of individual events (which will be the majority) should
grow approximately like ${\cal E}$, the number of clusters grows at a faster rate
$\overline{N}_m\propto {\cal E}^\alpha$. Of course the total number of events is expected
to grow strictly proportionally to the exposure, as  results 
from Eq.~(\ref{poisson}), since $\overline n\equiv \sum m \overline N_m=F_{tot}
{\cal E}$, with $F_{tot}$ the total flux. The fraction of events which are
in multiplets (neglecting here chance coincidences) grows hence, as long as
 ${\cal E}\ll F_{min}^{-1}$, as
${\cal E}^{\alpha-1}$.

\begin{figure}[t]
\centerline{{\epsfig{width=4.in,file=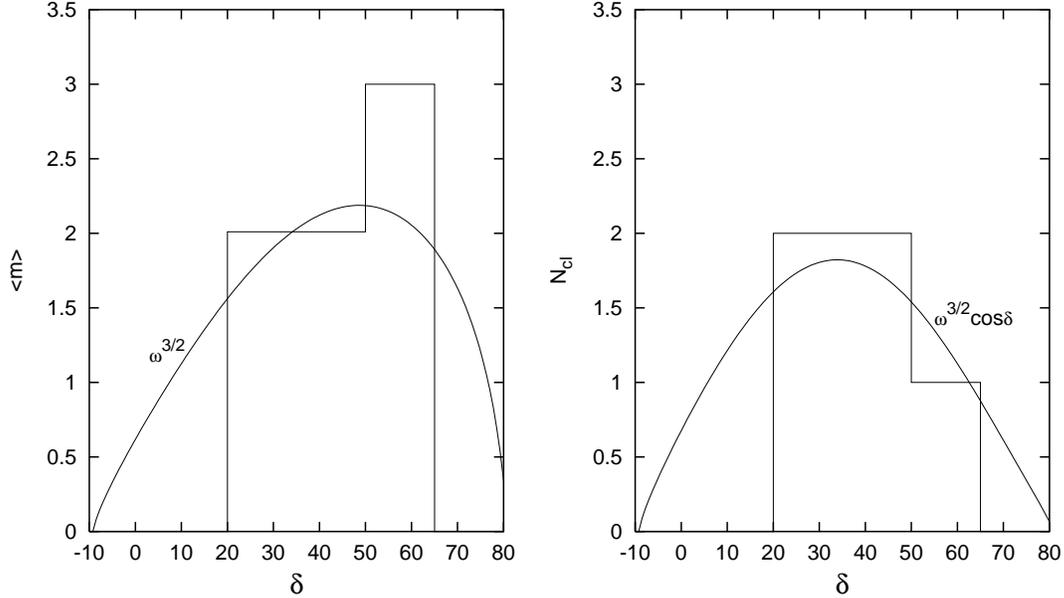,angle=-90}}}
\caption{Distribution in declination of cluster multiplicity (left panel) 
and number of clusters (right panel) in AGASA data (histograms) 
compared to that expected from an isotropic distribution of 
stationary sources ($\alpha=3/2$, see text), 
for which the overall normalization adopted is just arbitrary 
(solid curves).} 
\label{histo2}
\end{figure}

It is also interesting to compute the ratio of expected number of clusters
with different multiplicities,
\begin{equation}
{\overline N_l\over \overline N_m}\simeq {m !\over l !}
{\Gamma(l-\alpha)\over \Gamma(m-\alpha)}\ \ \ \ (l,m\ge 2),
\end{equation} 
and in particular for $l=m+1$ one obtains
\begin{equation}
{\overline N_{m+1}\over \overline N_{m}}\simeq {(m -\alpha)\over (m+1)}
\ \ \ \ (m\ge 2).
\end{equation} 
Another related result is that
\begin{equation}
{\sum_{m=3}^\infty\overline N_m\over \overline N_2}\simeq 
\sum_{m=3}^\infty \prod_{l=3}^m {(l-1-\alpha)\over l}.
\label{ratio}
\end{equation} 
Notice that these ratios are independent of the total number of events $\overline n$,
contrary to the case of chance coincidences. On the other hand, the dependence on the
distribution of the source fluxes is rather noticeable,
for instance, for $\alpha=3/2$ the ratio in eq. (\ref{ratio}) is 0.33,
while for $\alpha=5/4$ it becomes instead 0.60.

All these statistical quantities do not depend on whether the angular 
distribution of sources in the sky is isotropic or structured, as long 
as the average radial distribution does not differ much from uniformity.
Of course fluctuations would affect these average behaviours, and they
could come from a spread in the source luminosities, which could be either
intrinsic or caused by magnetic lensing effects \cite{ha02}  (this could
typically produce some clusters of larger than expected multiplicities),
from fluctuations in the radial source density distribution, from small 
statistics, etc.. 

Regarding the effects of varying exposures, it is clear that in the 
regions of larger exposures the sources are expected to yield clusters of
larger multiplicities on average, so that one expects that with 
large statistics, an uniform distribution of sources would lead to an 
average multiplicity varying as $\omega(\delta)^\alpha$.
On the other hand, the number of clusters (with multiplicities $m\geq 2$),
i.e. $N_{cl}=\sum_{m\geq 2}\overline{N}_m$, should follow the distribution 
d$N_{cl}\propto \omega(\delta)^\alpha\cos\delta$d$\delta$, and although the 
data are still scarce, the observed clusters are consistent with these 
behaviours, as illustrated in Figure~\ref{histo2}.
Notice that the total number of events (in clusters or not) should follow
d$n\propto \omega(\delta)\cos\delta$d$\delta$ if the source 
distribution is isotropic, so that a deviation from this behaviour
would be a clear signal of anisotropies, no matter what is the 
underlying reason for the clustering observed (i.e. chance coincidences vs. 
strong sources).

\section{Nearest neighbour distribution}

Up to now we have discussed in detail the properties of the distribution of
multiplets in the different possible clustering scenarios. One of the main 
criticisms to
the statistics discussed has been the dependence on the angular binning of the
data, as the results depend on its choice. We will discuss now another
statistics to analyse small scale clustering, the frequency
distribution of the angular distance to the nearest neighbours of each event,
which avoids this problem. Each event has a first, a second, and so on,
nearest neighbour event, and the idea is to look at the distribution of their
distances.  This statistics has some advantages over the autocorrelation
function of pairs when studying non-linear clustering, as it depends on 
the correlation functions of all orders. This is because the probability of finding
no events at a distance smaller than a given angle depends on
the correlation functions of all orders. 
This method has been used for example by Bahcall and Soneira to study
the distribution of stars and the fraction of binaries and triples \cite{ba81}.

\begin{figure}[t]
\centerline{{\epsfig{width=4.5in,file=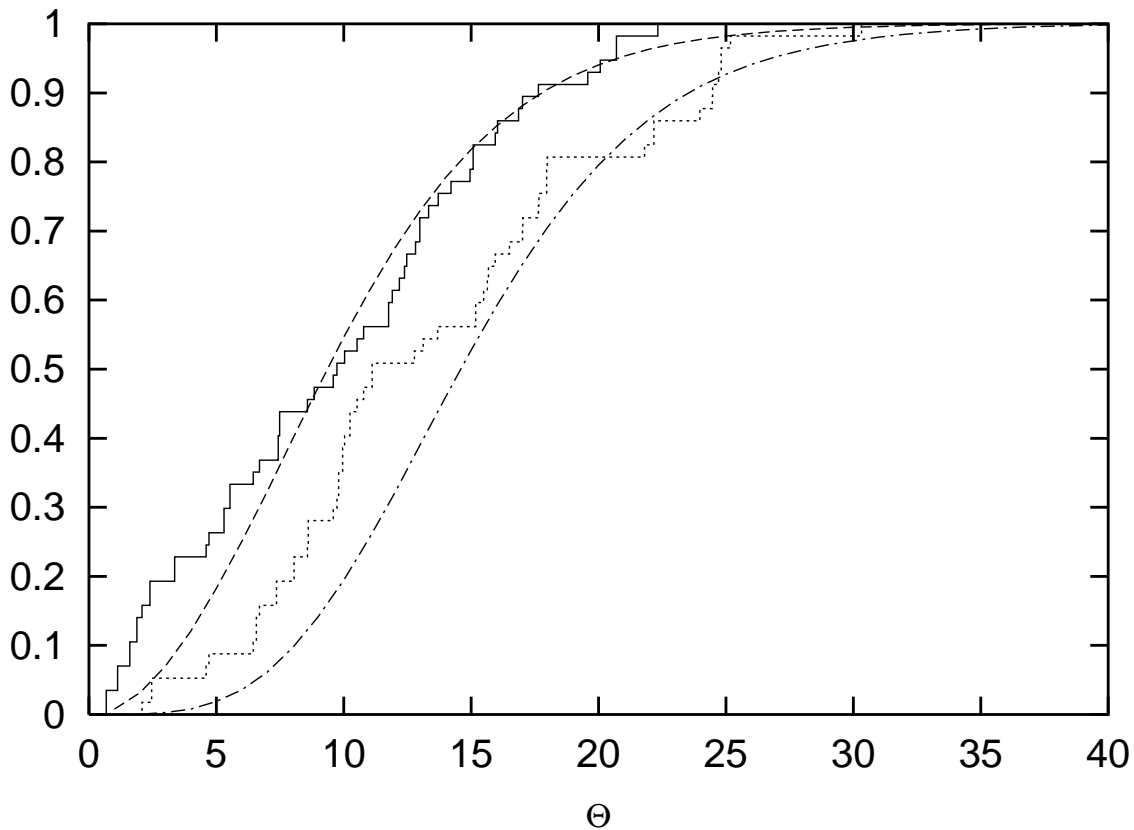,angle=-90}}}
\caption{Cumulative distribution of the distance to the first (solid histogram)
and second (dotted histogram) neighbour of the 57 AGASA events above $4\times 
10^{19}$ eV. The dashed and dashed-dotted curves are the mean distributions for 
10000 random realizations of 57 events with AGASA exposure.} 
\label{nn}
\end{figure}

The cumulative distributions of the distance to the first and second neighbour
for the 57 events of AGASA as a function of the angular distance are shown in
the solid histogram of Figure~\ref{nn}. 
The expected distribution for a sample of random
points in the sphere has been computed by Scott and Stout \cite{sc89}. 
However, in order
to compare with the distribution of the AGASA data we have to take into acount
the incomplete sky coverage and inhomogeneous exposure. Thus, we used instead
the
results of 10000 simulations of 57 random events with the AGASA exposure, the
mean of which is shown in Figure~\ref{nn} (dashed line). 
The AGASA distribution shows an excess
of first neighbours at small angles, with a maximal departure at $2.5^\circ$,
with the excess 
disappearing at angles slightly larger than $5^\circ$. This excess is clearly
related to the excess of doublets and triplets at small angles discussed
before. The cumulative histogram of the second neighbours (dotted histogram) 
also shows an
extended excess with respect to random expectations (dash-dotted line)
from angles around $2.5^\circ$ up to $15^\circ$, with a maximal deviation at
$11^\circ$. 

In order to have a measure of the significance of the first and
second neighbour excesses, the natural choice would be to perform 
 a Kolmogorov-Smirnov test. However, a necessary
condition for this test is that the data be all independent, and the nearest
neighbour separations between events are not independent variables (since
often the nearest neighbour of an event has as nearest neighbour that event).
Thus, we determine the significance of a given deviation $D$ of the cumulative
histogram from that expected from a random sample of the same number of events
using Montecarlo simulations and look for the fraction of the simulations
with a deviation from the mean larger than the observed $D$. 
The first neighbour
distribution shows a maximum deviation from the mean of the random simulations
$D=0.146$, with  $6\%$ of the
simulations having larger deviations at any angle, while the second neighbours 
curve maximum
deviation is $D=0.246$, with $0.8\%$ of the simulations showing larger
deviations. 

We have also checked the deviation of the mean angular separation to the first
and second neighbour of the data from that expected for a random sample. While
the AGASA data have first neighbour mean distance $\langle \theta_1
\rangle_{AG}=9.66^\circ$, the mean for a random sample is $\langle \theta_1
\rangle_{ran}=10.18^\circ$, with $25\%$ of the simulations having values
smaller than the AGASA one. For the second neighbours we find  $\langle
\theta_2 \rangle_{AG}=13.57^\circ$, while  $\langle \theta_2
\rangle_{ran}=15.35^\circ$, with only $2.3\%$ of the simulations having smaller
than AGASA's mean value. 

\begin{figure}[t]
\centerline{{\epsfig{width=4.5in,file=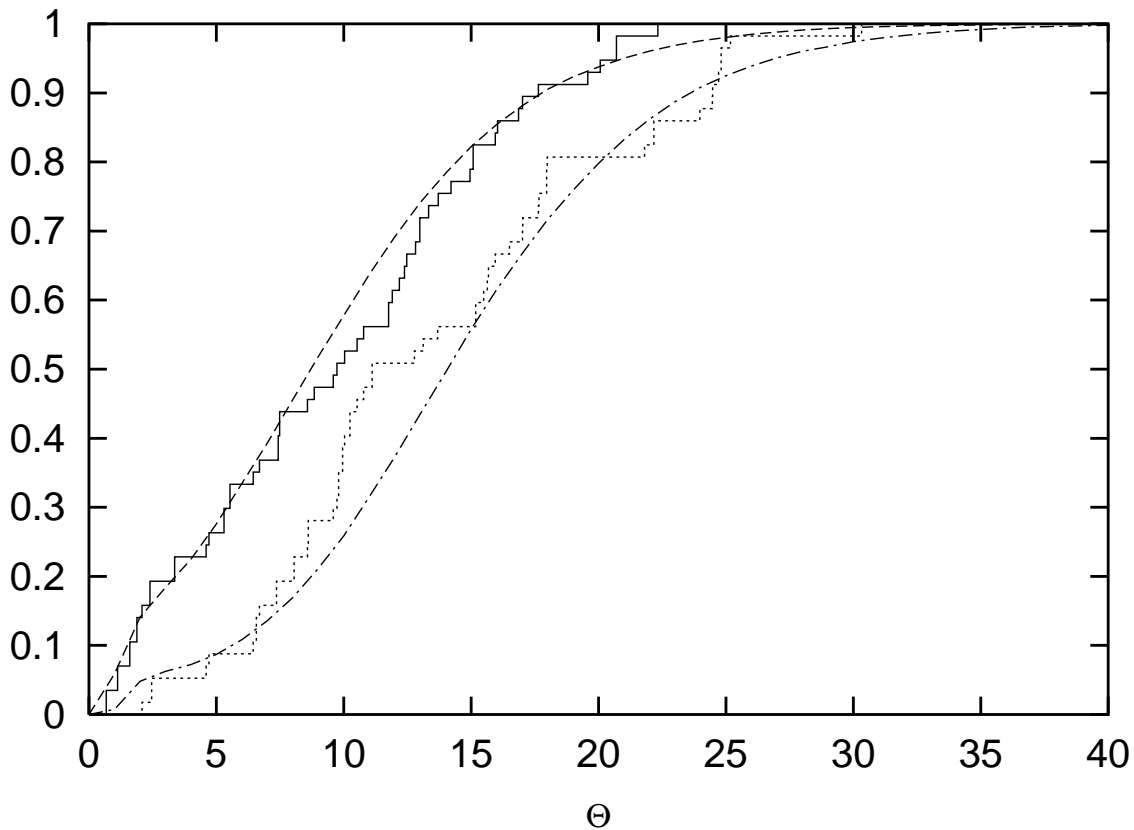,angle=-90}}}
\caption{Same as Figure~\ref{nn}, but now the simulations have 
53 events randomly thrown with AGASA exposure, and the remaining 4 events are
forced to form one triplet and two doublets within $2.5^\circ$ of the rest.} 
\label{nncl}
\end{figure}

Let us note that the maximum deviation of the second neighbour distance
cumulative distribution, that takes place at $\theta \simeq 11^\circ$, reflects
the clustering at all smaller scales. In particular, the effect of the triplet
is clearly seen in the rise of the histogram at $2.5^\circ$, which produces a
shift in the whole curve. Also the excess of doublets at small angles is
expected to affect the second neighbours distance distribution: the events
entering in the doublets will have their second neighbours at the mean
distance at which singlets have their first neighbours. Then, we may ask if
the whole deviation of the second neighbours at $11^\circ$ can be explained as
a result of the excess of small scale clustering (triplet and doublets). To
answer this, we performed a similar analysis, comparing the data with 10000
simulations of 57 events, 53 of which are randomly thrown with the AGASA
exposure, and the rest are forced to form a triplet and two doublets within
$2.5^\circ$ with some of the other events (chance alignments of random points
are expected to produce 1.5 additional doublets on average at $2.5^\circ$). 
The resulting cumulative
distributions for the first and second neighbours distance are shown in 
Figure~\ref{nncl}. The first neighbours distribution of the AGASA 
data are now in good agreement
with the result of the simulations, as well as the small angle piece of the
second neighbours distribution. There remains however a deviation of this 
distribution
with a maximum distance $D=0.187$ at $11^\circ$, with $7.6\%$ of the
simulations showing larger deviations. The mean distance to the second
neighbour of the simulation is now reduced to $\langle \theta_2\rangle_{ran}=
14.5^\circ$, with $15\%$ of the simulations showing values smaller than the
AGASA's one. While the significance of the deviation
is reduced by forcing the small scale clustering in the simulations, it
remains however a hint of extra clustering at angular scales around
$10^\circ$ that may be due to point source events that have been deflected by
magnetic fields, or to structure in the spatial distribution of sources.

\section{Discussion}

We have seen that the statistics of clusters of events with different multiplicities
is a very useful tool that may help characterise the nature of UHECR sources.
We have analysed the way in which the relative number of clusters with different
multiplicities, their change with increasing exposure, and their 
distribution with declination
can help discriminate among alternative source scenarios. 
In particular, one of the most striking differences between the individual 
sources and chance coincidences scenarios for clustering is the expected 
hierarchy
of the number of clusters with different multiplicities. For example, the
ratio between the number of triplets or higher multiplets and doublets
is $\sim 1/3$ for the steady source model (or 0.6 for the bursting one), 
while it is only $\sim n/3N$ for the chance coincidences, 
where $N\sim 10^3(2.5^\circ/\theta)^2$.
In case the clusters are dominated by individual sources, we 
see that with significant statistics this may even give a handle to discriminate 
between bursting and steady source scenarios. Moreover, the evolution of 
this ratio (which is proportional to the exposure for chance coincidences,
 while independent of it for individual sources) will give a further tool to 
discriminate among scenarios.
On the other hand, the relative fraction of events in clusters (with respect 
to the unclustered ones) is only $\sim n/2N$ for the chance coincidences of 
an isotropic distribution, but could be significantly enhanced if the 
distribution of sources is anisotropic. This should be however testable 
by studying the overall event distribution once larger statistics are achieved.
In the case of clusters produced by individual sources, this fraction 
depends instead sensitively on the amount of very faint sources present.

We have applied our analysis
to the published set of 57 AGASA events above $4\times 10^{19}$ eV just 
for illustrative
purposes, since it is clearly insufficient to draw definitive conclusions. 
Nevertheless, our results indicate that a modest increase in the number 
of events may already give
much stronger hints about the source properties. In this
respect, even the release of the latest AGASA data, which represents a 50\%
increase with respect to the published data, could have some impact. For 
instance, it would be natural to expect that the triplet becomes a quadruplet, and
some doublets become triplets, if they are due to individual sources, while 
this would be very unlikely if they arose from chance coincidences.

We have also applied an alternative statistical tool to analyse UHECR 
clustering: the nearest neighbour distribution. It has the advantage 
over the pair autocorrelation method of being less sensitive to binning, and
that it depends not just on the pair correlations but on correlations to all 
orders \cite{wh79}. Applied to AGASA data, it gives an alternative measure
of the small angular scale clustering, and also points out to an excess of clustering
with respect to random expectations which persists up to $15^\circ$, that manifests
in the second neighbours cumulative distribution (Figure~\ref{nn}), and which is also
hinted by the defect of doublets and excess of triplets at scales up to $15^\circ$
in Figure~\ref{pdt}. Angular spread of events in a cluster due to deflections by
magnetic fields, and structure in the spatial distribution of sources, are among
conceivable causes of such signal, which should be looked upon with better statistics.

\section*{Acknowledgments}
Work supported by ANPCyT, CONICET, Fundaci\'on Antorchas and the Guggenheim Foundation.

\end{document}